\documentstyle[titlepage,fleqn,12pt]{article}
\begin{document}

\setcounter{page}{0}

\begin{center}
{\large \bf Effects of long-range coupling on aggregation }
\end{center}

\begin{center}
 E. Canessa$^{\dag}$$(*)$  and Wei Wang$^{\dag\ddag}$\\

\vspace{1.2cm}

$^{\dag}$International Centre for Theoretical Physics, Trieste, Italy

\vspace{0.5cm}

$^{\ddag}$Physics Department, Nanjing University, People's Republic of China
\end{center}

\vspace{12cm}

\vspace{1cm}

-------------------------------

$^{(*)}$ {\footnotesize Reprint requests to:
canessae@ictp.trieste.it}

\baselineskip=24pt
\parskip=0pt

\newpage
\begin{center}
{\bf Abstract }
\end{center}

Numerical simulations of a 2D biharmonic equation $\nabla^{4}u =0$
show that a transition from dense to multibranched growth is {\em only}
a consequence of long-range coupling on the pattern formation of fractal
aggregates.

\newpage

It has been recently shown that screening, due to free charges, strongly
diversifies the patterns that grow in electrostatic fields \cite{Lou92,Cas93},
(see also \cite{Wan93}).
By solving the Poisson equation, it has been found a transition from dense to
multibranched growth which depends on the potentials $\phi^{o}$ and $\phi^{i}$
at two boundaries, the distance between them, $L-\ell$, and an inverse
screening length $\lambda$.  In this note we illustrate an extention of this
problem based on the biharmonic (BH) equation $\nabla^{4}u =0$ in 2D isotropic
defect-free media.  We prove here that the transition from dense to
multibranched growth is {\em only} a consequence of long-range coupling
between displacements on pattern formation and that in the present model the
transition appears when the velocity on the growing
surfaces present a minimum as also occurs in Poisson growth.

These new results are important because of the physical relevance of the BH
equation as follows from the well-known Kuramoto-Sivashinsky (KS) equation
that models pattern formation in different physical contexts, such as chemical
reaction-diffusion systems and cellular gas flames in the presence of external
stabilizing factors \cite{Sne92,Ma93}.  The BH equation also appears,
{\em e.g.}, when describing the deflection of a thin plate subjected to
uniform loading over its surface with fixed edges or within the steady,
slow 2D motion of a viscous fluid \cite{Mel90}.

Herein we consider the simplest version of the KS equation, {\em i.e.} we
assume static solutions.  This allows us to include long-range coupling
through the discretization of the BH equation on lattice sites involving
values of $u$ at {\em thirteen} mesh points.  This is the {\em crucial}
difference with respect to Laplacian and Poisson models in which iterative
procedures are carried out around ({\em four}) next-nearest neighbours (nn)
only.

Figure 1 shows the final stage of a BH pattern displaying features, in
circular geometry, of a transition from dense to multibranched growth
when attaching one particle at each step.
Below the transition point $r_{\ell} \sim 0.6\; L$, the fractal dimension
approaches the value for Laplacian growth within error bars.
To generate this BH fractal pattern we have set the derivative boundary
condition (DBC), that is necessary along the $r$-direction, equal to zero
and the growth probability $P$ proportional to $\nabla^{2} u$,
(corresponding to the potential in \cite{Lou92}).
Above $r_{\ell}$ this figure is a demostration that long-range coupling is the
most relevant aspect for this phenomenon rather than screening as suggested
by Louis {\em et al}.

For planar growth we use periodic boundary conditions in the $x$-direction
and estimate the DBC along the $y$-direction from the analytical
solution $3u^{o}/L$, by fixing $u^{i}=0$ and
(rescaled) $\phi^{o}\equiv \nabla^{2}u\mid_{y=L}\approx 6u^{o}/L^{2}$.

In Fig.2 we plot results for the grow velocity $v$ along the $y$-direction,
which we relate to the averaged value of $P$, {\em e.g.}, equal to
$\nabla^{2}u$ (curve A), to include nn sites, or to the
displacement $u$ (curve B). In curve C $v$ is set proportional to the
field $\mid \phi_{i,j}-\phi^{i}\mid$ following \cite{Lou92}.
In this geometry we find that the transition appears when $v$
on the growing surfaces present a minimum as also occurs in Poisson growth
(PG: curve C).  However, the BH patterns in the dense region
(not shown) are not that denser as in Refs.\cite{Lou92,Cas93}. The
reason for this is that for PG the transition occurs at much smaller $v$
than for BH growth, hence, a Eden-like pattern can be generated.
Above this transition multibranched fractals appear in both models, but
for BH growth the transition line depends on the system size.
It is remarkable that the three curves in Fig.2 present parallel slopes
above and below their respective transition points.

The transition obtained by numerically solving the BH part of the KS
equation might not be necessarily similar to the Hecker transition.
But, the KS equation can be transform to look somewhat like the Navier-Stokes
equation for a potential flow with negative viscosity which may be somehow
related to the recent analysis in Ref.\cite{Fleu92} concerning electrochemical
deposition.

\newpage

\section*{Figure captions}

\begin{itemize}

\item {\bf Fig.1}:
Biharmonic pattern diplaying the transition.

\item {\bf Fig.2}:
Growth velocity $v$ along the $y$-direction proportional to
$\mid \nabla^{2}u \mid$ (A) or displacement $\mid u \mid$ (B).
(C) is for Poisson growth \cite{Lou92}.

\end{itemize}

\end{document}